\documentclass[aps,prb,twocolumn,showpacs,floatfix]{revtex4}

\usepackage{dcolumn}
\usepackage[english]{babel}
\usepackage{amsfonts,amsmath,amssymb,bm}

\begin{document}

\title{DYNAMIC LOCALIZATION EFFECTS IN L-RING CIRCUIT}
\author{C.MICU (a), E. PAPP (b) , L. AUR (b)}
\email{erhardt_papp_2005@yahoo.com} \affiliation{ (a) Physics
Department, North University of Baia Mare, RO-430122, (b) Department
of Theoretical Physics, West University of Timisoara, RO-300323}
\date{\today}

\begin{abstract}

 Using suitable magnetic flux operators established in
terms of discrete derivatives leads to quantum-mechanical
descriptions of LC-circuits with an external time dependent periodic
voltage. This leads to second order discrete Schrodinger equations
provided by discretization conditions of the electric charge.
Neglecting the capacitance leads to a simplified description of the
L-ring circuit threaded by a related time dependent magnetic flux.
The equivalence with electrons moving on one dimensional (1D)
lattices under the influence of time dependent electric fields can
then be readily established. This opens the way to derive dynamic
localization conditions serving to applications in several areas,
like the time dependent electron transport in quantum wires or the
generation of higher harmonics by 1D conductors. Such conditions,
which can be viewed as an exact generalization of the ones derived
before by Dunlap and Kenkre [Phys. Rev. B 34, 3625(1986)], proceed
in terms of zero values of time averages of related persistent
currents over one period.

\textbf{Keywords}: Charge discretization; Quantum L-ring circuits;
Persistent currents; Dynamic localization conditions; Quantum wires

\end{abstract}

\maketitle

\section{Introduction}

Quantum circuits with charge discreteness have been the focus considerable
interest during the last decade [1-5]. It has been realized that quantum
circuits are able to provide useful developments in the field of nanodevices,
transmission lines as well as of molecular electronic circuits [6-8]. The
chargediscreteness referred to above gets incorporated in the eigenvalue equation%

\begin{equation}
Q|n>=nq_{e}|n> \tag{1}%
\end{equation}
where $n$ is an integer and where $Q$ denotes the Hermitian operator of the
electric charge. The elementary charge is denoted by\ $q_{e}$. The common
choice is to put $q_{e}=e$, where $e$ is the charge of the electron. The
charge eigenfunctions are orthonormalized , in which case the time dependent
wavefunction describing the quantum circuit can be expressed as%

\begin{equation}
|\psi(t)>=%
{\displaystyle\sum\limits_{n}}
C_{n}(t)|n> \tag{2}%
\end{equation}
where $n\epsilon\left(  -\infty,\infty\right)  $ We have also to keep in mind
that usual derivatives have to be replaced by discrete ones when dealing with
$n$ - dependent functions. Proceeding in this manner leads to a discrete
Schrodinger equation, as shown before [1-3]. On the other hand, the influence
of the capacitance can be disregarded, which results in a discrete
Schr\"{o}dinger-equation concerning the L-ring circuit. The interesting point
is that this latter equation is equivalent, under suitable matching
conditions, to the one describing the 1D conductor, i.e. to the electron on
the 1D lattice under the influence of a time dependent electric field. This
opens the way to a general derivation of dynamic localization conditions
(DLC's) for periodic modulations of the voltage.

\section{Preliminaries and notations}

One starts from the classical description of Kirchhoff's law for the
LC-circuit with a voltage source in terms of Hamilton's equations. This
amounts to consider a classical Hamiltonian like $H\left(  \Phi/c,Q\right)  $,
where $\Phi$ denotes the magnetic flux. Accordingly $Q$, and $\Phi/c$ play the
role of the coordinate and momentum, respectively. The next step is to perform
the quantization of canonically conjugated variables. For this purpose, the
usual canonical commutation relations could be applied [9, 10]. However, we
have to account for the discreteness of the electric charge, which means that
the appropriate quantum mechanical description of the magnetic flux should
proceed in terms of discrete derivatives [2, 3]. Under such conditions one
gets faced with a deformation of the Heisenberg algebra emphasized usually.
This results in the Hamiltonian%

\begin{equation}
H\left(  \frac{\Phi}{c},Q\right)  =\frac{1}{2Lc^{2}}\Phi^{+}\Phi+\frac{Q^{2}%
}{2C}-QV_{s}(t) \tag{3}%
\end{equation}
where $V_{s}(t)$ denotes the external time dependent voltage. Now the magnetic
flux is described by the non-Hermitian operator%

\begin{equation}
\Phi=-i\frac{\hbar c}{q_{e}}\Delta\tag{4}%
\end{equation}
so that $\Phi^{+}=-i(\hbar c/q_{e})\Delta$ , where the \textquotedblleft%
$+$\textquotedblright\ superscript stands for the Hermitian conjugation. The
right and left-hand discrete derivatives displayed above act as $\Delta
f(n)=f(n+1)-f(n)$ and $\nabla f(n)=f(n)-f(n-1)$, respectively. The present
quantization condition is then given by the deformed relationship%

\begin{equation}
\left[  \Phi,Q\right]  =i\hbar c\left(  1+i\frac{q_{e}}{\hbar c}\Phi\right)
\tag{5}%
\end{equation}
which differs from the usual canonical commutation relation by virtue of the
deformation rule $1\rightarrow1+\left(  iq_{e}\Phi\right)  /\left(  \hbar
c\right)  $, such as displayed by the r.h.s. in (5). Resorting to (2), then
yields the discrete time dependent Schr\"{o}dinger-equation [1]%

\begin{equation}
-\frac{\hbar^{2}}{2Lq_{e}^{2}}\left(  C_{n+1}(t)+C_{n-1}(t)\right)
+MC_{n}(t)=i\hbar\frac{\partial}{\partial t}C_{n}(t) \tag{6}%
\end{equation}
in which the $\left(  \hbar^{2}c^{2}\right)  /\left(  Lq_{e}^{2}\right)
$-term in (6) can be ruled out by virtue of the gauge transformation%

\begin{equation}
C_{n}(t)\longrightarrow C_{n}(t)e^{-i\frac{\hbar c^{2}}{Lq_{e}^{2}}} \tag{7}%
\end{equation}
where $M=\left(  \frac{q_{e}^{2}}{2C}n^{2}-nq_{e}V_{s}(t)+\frac{\hbar^{2}%
c^{2}}{Lq_{e}^{2}}\right)  $

\section{The duality between the L-ring circuit and the electron on the 1D
lattice}

\qquad\qquad Neglecting the capacitance and applying the $C\longrightarrow
\infty$ - limit, the presence of the harmonic oscillator term in (6) is ruled
out. This results in an explicit discrete Schr\"{o}dinger equation for the
L-ring circuit only. On the other hand the electron on the 1D lattice under
the influence of an external time dependent electric field like $E_{e}%
(t)=(\hbar E_{F}f(t))/(ea)$, where \ $a$ denotes the lattice spacing, is
characterized by the evolution equation [11]%

\begin{equation}
\hbar V\left(  C_{n+1}(t)+C_{n-1}(t)\right)  -\hbar E_{F}nf(t)C_{n}%
(t)=i\hbar\frac{\partial}{\partial t}C_{n}(t) \tag{8}%
\end{equation}

Keeping in mind that $C\longrightarrow\infty$, one sees that (6) reproduces
precisely (8) in terms of the matching conditions%
\begin{equation}
V_{s}(t)=\frac{\hbar}{q_{e}}E_{F}f(t)=-\frac{d}{cdt}\Phi_{e}(t) \tag{9}%
\end{equation}
and $\hbar V=-\hbar^{2}/(2Lq_{e}^{2})$ . One remarks, of course, that (9)
incorporates Faraday`s law. Accordingly, there is a duality between the L-ring
circuit and the electron on the 1D-lattice, which is useful for the study of
dynamic localization effects [11].

\section{Dynamic localization effects}

Let us consider a sinusoidal modulation of the magnetic flux like%

\begin{equation}
\Phi_{e}(t)=\Phi_{0}\sin(wt) \tag{10}%
\end{equation}

This yields a time dependent electric field modulated by the characteristic function%

\begin{equation}
f(t)=\cos(wt) \tag{11}%
\end{equation}

so that the field amplitude is given by%

\begin{equation}
E_{F}=-\frac{q_{e}}{\hbar c}\Phi_{0}w \tag{12}%
\end{equation}

Now we have to remember that there is a periodic return of the electron to the
initially occupied site if the quotient $E_{F}/w$ is a root of the Bessel
function of order zero and of the first kind [11]:%

\begin{equation}
J_{0}\left(  \frac{E_{F}}{w}\right)  =0 \tag{13}%
\end{equation}

Accordingly, the mean square displacement (MSD)

\qquad%
\begin{equation}
<n^{2}>=%
{\displaystyle\sum\limits_{n=-\infty}^{n=\infty}}
|C_{n}(t)|^{2}n^{2} \tag{14}%
\end{equation}

Remains bounded in time, which is synonymous to the onset of the dynamic
localization. We then have to realize that this latter effect should also
concern the carriers of the discretized charge in the L-ring circuit. Indeed,
(13) can be rewritten equivalently as%

\begin{equation}
J_{0}\left(  \frac{q_{e}\Phi_{0}}{\hbar c}\right)  =0 \tag{15}%
\end{equation}

whereas the corresponding MSD has been established [3] in terms of velocity
autocorrelation functions [12]. Dynamic localization conditions like (13)
and/or (15) are of interest to the study of time dependent electron transport
through quantum wires [13, 14], of the generation of higher harmonics by 1D
conductors [15] as well as of other areas [16]. The dynamic localization has
been observed in the linear optical absorption spectra of quantum dot
superlattices [17]. Negative conductances observed in photon assisted
tunneling effects can also be invoked [18].

\section{Persistent currents}

In order to proceed further, let us remember that the persistent current
carried by a flux dependent energy level $E=E(k,\Phi_{e})$ at is [19]%

\begin{equation}
I_{k}(\Phi_{e})=-c\frac{\partial}{\partial\Phi_{e}}E(k,\Phi_{e}) \tag{16}%
\end{equation}

where the presence of the wavenumber $k$ accounts for the existence of energy
bands. We shall also assume that $f(t)$ in (9) is periodic in time with period
$T$. On the other hand, the energy dispersion law characterizing (8) reads%

\begin{equation}
E(k)=2\hbar V\cos(ka) \tag{17}%
\end{equation}
such as produced by inserting the free-field amplitude

\qquad\qquad%
\begin{equation}
C_{n}(t)=e^{-\frac{i}{\hbar}E(k)t}e^{ikan} \tag{18}%
\end{equation}
into the $E_{F}=0$\ form of (8). What then remains is to account for the
external time dependent electric field%

\begin{equation}
E_{e}(t)=-\frac{1}{c}\frac{\partial}{\partial t}A_{1}(t) \tag{19}%
\end{equation}
in terms of the minimal substitution%

\begin{equation}
\hbar k\longrightarrow\hbar k+\frac{e}{c}A_{1}(t) \tag{20}%
\end{equation}
in which%

\begin{equation}
A_{1}\left(  t\right)  =-\frac{\hbar c}{ea}E_{F}\eta\left(  t\right)  \tag{21}%
\end{equation}
and

\qquad%
\begin{equation}
\eta\left(  t\right)  =%
{\displaystyle\int\limits_{0}^{t}}
f\left(  s\right)  ds \tag{22}%
\end{equation}
Accordingly, (17) becomes

\qquad%
\begin{equation}
E(k)\longrightarrow E(k,\Phi_{e})=2\hbar V\cos\left(  ka+\frac{q_{e}}{\hbar
c}\Phi_{e}(t)\right)  \tag{23}%
\end{equation}
which yields the persistent current%

\begin{equation}
I_{k}\left(  \Phi_{e}\right)  =-\frac{\hbar}{q_{e}L}\sin\left(  ka+\frac
{q_{e}}{\hbar c}\Phi_{e}\right)  \tag{24}%
\end{equation}
by virtue of (16). Similar results have been written down before [3] by
resorting to the -wavenumber representation of (6).

\section{Dynamic localization conditions for arbitrary time periodic
modulations}

Performing the time average over one period of (24) yields the averaged current%

\begin{equation}
J_{k}(T)=-\frac{\hbar}{q_{e}LT}%
{\displaystyle\int\limits_{0}^{T}}
\sin\left(  ka+\frac{q}{\hbar c}\Phi_{e}(t)\right)  dt \tag{25}%
\end{equation}
which deserves further attention. Indeed, accounting for (11), one finds that
the DLC (13) gets reproduced whenever%

\begin{equation}
J_{k}(T)=0 \tag{26}%
\end{equation}
irrespective of $k$ . For this purpose we have to resort to the generating
function of Bessel functions of the first kind [20]%

\begin{equation}
e^{iz\sin\varphi}=%
{\displaystyle\sum\limits_{m=-\infty}^{m=+\infty}}
J_{m}(z)e^{im\varphi} \tag{27}%
\end{equation}
It is also clear that integrating (9) yields

\qquad%
\begin{equation}
\Phi_{e}\left(  t\right)  -\Phi_{e}\left(  0\right)  =-\frac{\hbar c}{q_{e}%
}E_{F}\eta(t) \tag{28}%
\end{equation}
so that gets specified in terms of and conversely.

The dc-ac electric field $E(t)=E_{0}+E_{1}\cos\left(  wt\right)  $\ for which%

\begin{equation}
f(t)=\frac{w_{B}}{E_{F}}+\cos(wt) \tag{29}%
\end{equation}
can also be readily discussed. Now one has $E_{F}=\left(  E_{1}ea\right)
/\hbar$ , while denotes the Bloch frequency. Inserting (29) into (26) yields
the DLC%

\begin{equation}
J_{n}\left(  \frac{E_{F}}{w}\right)  =0 \tag{30}%
\end{equation}
if $w_{B}=nw$ , where $n$\ is a positive integer, such as established before
within the quasi-energy description [21,22]. This latter equation can be
viewed as an exact $n\neq0$ counterpart of (13). Rational realizations of the
matching ratio $\left(  w_{B}/w\right)  $ require, however, a special
treatment [21, 22].

Other cases can be treated in a similar manner.

\section{Conclusions}

\qquad In this paper the quantum L-ring circuit threaded by a time dependent
magnetic flux has been discussed with a special emphasis on the equivalence
with the electron on the 1D lattice under the influence of a periodic time
dependent electric field. Proceeding in this manner leads to (26), which has
the meaning of an exact generalization of (13). Such results are useful for
applications in several areas already referred to above. Generalizations going
beyond the nearest neighbor description deserve further attention, too [23].

\qquad\qquad\textbf{Acknowledgment}

\bigskip

\qquad\qquad We are indebted o CNCSIS/Bucharest for financial support.

\bigskip

\qquad\qquad\textbf{References}

\bigskip

[1] Y. Li and B. Chen, Phys. Rev. B 53, 4027 (1996)

[2] T. Lu and Y. Q. Li, Mod. Phys. Lett. B 16, 975 (2002)

[3] B. Chen, X. Dai and R. Han, Phys. Lett. A 302, 325 (2002)

[4] B. Chen et al, Phys. Lett. A 335, 103 (2005)

[5] J. C. Flores, M. Bologna, K.J. Chandia and C. A. Utreras Diaz, Phys. Rev.
B 74, 193319 (2006)

[6] S. Ami and C. Joachim, Phys. Rev. B 65, 155419 (2002)

[7] A. Wacker, Phys. Rep. 357, 1 (2002).

[8] M. Lieberman (Ed.), Molecules as Components of Electronic Devices (Oxford
University Press, Oxford, 2003).

[9] W. H. Louisell, Quantum Statistical Properties of Radiation (John Willey,
New York, 1973)

[10] J. C. Flores and C. A. Utreras Diaz, Phys. Rev. B 66, 153410 (2002)

[11] D. H. Dunlap and V. M. Kenkre, Phys. Rev. B 34, 3625 (1986)

[12] V. M. Kenkre, R. K\"{u}hne and P. Reineker, Z. Phys. B 41, 177 (1981)

[13] T. Kwapinski, Phys. Rev. B 69, 153303 (2004)

[14] I. Tralle, Physica E 9, 275 (2001)

[15] K. A. Pronin, A. D. Bandrauk and A. A. Ovchinnikov, Phys. Rev. B 50, 3473 (1994)

[16] V. M. Kenkre and S. Raghavan, J. Opt. B: Quantum Semiclss. Opt. 2, 686 (2000)

[17] J. R. Madureira, P. A. Schulz and M. Z. Maialle, Phys. Rev. B 70, 033309 (2004)

[18] B. J. Keay, S. Zeuner, S. J. Allen Jr., K. D. Maranowski, A. C. Gossard,
U. Bhattacharya and M. J. Rodwell, Phys. Rev. Lett. 75, 4102 (1995)

[19] H. F. Cheung, Y. Gefen, E. K. Riedel and W. H. Shih, Phys. Rev. B 37,
6050 (1988)

[20] I. S. Gradshteyn and I. M. Ryzhik, Table of Integrals, Series and
Products ( Academic Press, New York, 1980)

[21] A. Z. Zhang, P. Zhang, D. Suqing, X. G. Zhao and J. Q. Liang, Phys. Lett.
A 292, 275 (2002)

[22] X. G. Zhao, R. Jahnke and Q. Niu, Phys. Lett. A 202, 297 (1995)

[23] M. A. Jivulescu and E. Papp, J. Phys. Condens. Matter 18, 6853 (2006)

\end{document}